\documentclass[superscriptaddress,aps,prb,showpacs,twocolumn]{revtex4-1}

\usepackage{amsmath}
\usepackage{graphicx}
\usepackage[ansinew]{inputenc}
\usepackage{color,ulem}
\usepackage{xcolor}

\newcommand{\gb}{$\bar{\Gamma}$}
\newcommand{\kb}{$\bar{K}$}
\newcommand{\mb}{$\bar{M}$}

\newcommand{\bise}{Bi$_2$Se$_3$}
\newcommand{\bite}{Bi$_2$Te$_3$}

\begin{document}

\title{Intra- and Interband Electron Scattering in the Complex Hybrid Topological Insulator Bismuth Bilayer on Bi$_2$Se$_3$}

\author{A.~Eich}
\thanks{These authors have contributed equally to this work.}
\affiliation{Institute for Applied Physics, Universit\"{a}t Hamburg, D-20355 Hamburg, Germany}
\author{M.~Michiardi}
\thanks{These authors have contributed equally to this work.}
\affiliation{Department of Physics and Astronomy, Interdisciplinary Nanoscience Center (iNANO), Aarhus University, DK-8000 Aarhus C, Denmark}
\author{G.~Bihlmayer}
\affiliation{Institute for Advanced Simulation and Peter Gr\"{u}nberg Institut 1, Forschungszentrum J\"{u}lich, 52425 J\"{u}lich, Germany}
\author{X.-G. Zhu}
\affiliation{Department of Physics and Astronomy, Interdisciplinary Nanoscience Center (iNANO), Aarhus University, DK-8000 Aarhus C, Denmark}
\author{J.-L. Mi}
\affiliation{Center for Materials Crystallography, Department of Chemistry, Interdisciplinary Nanoscience Center (iNANO), Aarhus University, DK 8000 Aarhus C, Denmark}
\author{Bo B. Iversen}
\affiliation{Center for Materials Crystallography, Department of Chemistry, Interdisciplinary Nanoscience Center (iNANO), Aarhus University, DK 8000 Aarhus C, Denmark}
\author{R.~Wiesendanger}
\affiliation{Institute for Applied Physics, Universit\"{a}t Hamburg, D-20355 Hamburg, Germany}
\author{Ph.~Hofmann}
\affiliation{Department of Physics and Astronomy, Interdisciplinary Nanoscience Center (iNANO), Aarhus University, DK-8000 Aarhus C, Denmark}
\author{A.~A.~Khajetoorians}
\email[Corresponding author: ]{akhajeto@physnet.uni-hamburg.de}
\affiliation{Institute for Applied Physics, Universit\"{a}t Hamburg, D-20355 Hamburg, Germany}
\author{J.~Wiebe}
\affiliation{Institute for Applied Physics, Universit\"{a}t Hamburg, D-20355 Hamburg, Germany}
\date{\today}

\begin{abstract}
The band structure, intra- and interband scattering processes of the electrons at the surface of a bismuth-bilayer on \bise~have been experimentally investigated by low-temperature Fourier-transform scanning tunneling spectroscopy. The observed complex quasiparticle interference patterns are compared to a simulation based on the spin-dependent joint density of states approach using the surface-localized spectral function calculated from first principles as the only input. Thereby, the origin of the quasiparticle interferences can be traced back to intraband scattering in the bismuth bilayer valence band and \bise~conduction band, and to interband scattering between the two-dimensional topological state and the bismuth-bilayer valence band. The investigation reveals that the bilayer band gap, which is predicted to host one-dimensional topological states at the edges of the bilayer, is pushed several hundred milli-electronvolts above the Fermi level. This result is rationalized by an electron transfer from the bilayer to \bise~which also leads to a two-dimensional electron state in the \bise~conduction band with a strong Rashba spin-splitting, coexisting with the topological state and bilayer valence band.
\end{abstract}



\maketitle

\section{Introduction}

Since the seminal work of Murakami~\cite{Murakami:2006b}, a bilayer (BL) of bismuth in the (111) orientation is theoretically considered~\cite{Wada:2011} as one of the few two-dimensional (2D) electron systems, besides HgTe/(Hg,Cd)Te quantum wells~\cite{Konig:2007}, which should show the quantum spin Hall effect, and enter the quantum anomalous spin Hall phase by magnetic doping~\cite{Zhang:2012b}. The experimental realization of the Bi BL as a 2D topological insulator (TI) in a solid state environment remains challenging because of the need to find a suitable substrate which (i) enables epitaxial growth of the Bi BL and (ii) sufficiently decouples the electronic states of the BL from the substrate electrons in order to protect the predicted 1D topological state (TS) at the edges of the BL. Recently, it was experimentally shown that an epitaxial BL can be grown on the 3D TIs \bite~\cite{Hirahara:2011, Miao:2013c, Wang:2013, Coelho:2013} and \bise~\cite{Miao:2013c, Wang:2013}. Scanning tunneling spectroscopy (STS) showed indications for the survival of the 1D TS at the edges of BL islands grown on Te terminated surfaces of binary ~\cite{Yang:2012} and ternary \cite{Kim:2014} Bi-chalcogenides. However, the edge state's energetic position is still controversial~\cite{Yang:2012,Kim:2014}, partly because the band structure is complicated by the complex coexistence of the 1D Bi BL edge state and the 2D TS of the hybrid Bi BL-substrate system~\cite{Hirahara:2011, Miao:2013c, Wang:2013}.

Here, we experimentally investigate the band structure and scattering of the electron system of a Bi BL grown on \bise~by observation of the quasi-particle interference (QPI)~\cite{Crommie:1993,Hasegawa:1993} via Fourier-transform STS (FT-STS)~\cite{Hofmann:1997, Sprunger:1997, Petersen:1998a, Simon2011}. The FT-STS images are interpreted with the help of the joint-density of states (JDOS) approach, which has been used successfully to help the interpretation of FT-STS images for high temperature superconductors \cite{Hoffman:2002,McElroy2006, Haenke2012}, quantum-Hall electron phases in semiconductors~\cite{Hashimoto:2012}, and, more recently, in the field of TIs~\cite{Roushan2009,Okada:2011,Fang2013}. The JDOS method is usually applied with a spectral function $A(E,\textbf{k})$ determined from angle resolved photoemission spectroscopy (ARPES). Deducing $A$ from ARPES presents three major inconveniences: (i) A limitation to the occupied states. (ii) Very high quality ARPES data, which is needed in order to disentangle the spectral features, is often hindered by the challenge of obtaining a high quality crystal surface over micro meter distances. (iii) In systems with multiple bands in a small energy and momentum range, the interpretation of ARPES data can be problematic due to matrix element effects.
Instead, here, a semi-theoretical approach is developed, determining $A$ from a density functional theory (DFT) calculation of the band structure as input for the JDOS simulation, where the structural parameters have been optimized in order to fit the ARPES-measured band structure. The JDOS approach, in comparison to the more rigorous stationary phase approach~\cite{Liu:2012} or transfer matrix method~\cite{Wang:2003}, bears sufficient simplicity in order to easily investigate the effects of different spin-dependent scattering matrices $M$ on the simulated QPI patterns. On the other hand, the "on-shell" assumption used in the JDOS approach has been shown to be a sufficiently accurate simplification by comparison between JDOS and transfer matrix results for high-temperature superconductors~\cite{Wang:2003}.

The first objective of this paper is to demonstrate that the spin-dependent JDOS method using the DFT-calculated surface-localized spectral function applied here, can serve to interpret the complex FT-STS measured QPI patterns of Bi BL on \bise, involving up to three scattering vectors at a given energy. The second objective is the experimental investigation of the band structure and electron scattering in this TI. It is shown that the 2D TS of this system coexists over a large energy range with Rashba spin-split quantum well states in the conduction band (CB) and valence band (VB) of the substrate. The band gap of the Bi BL which hosts the 1D edge state is relocated several hundred meV above the Fermi level by charge transfer, and thus may be accessible to transport experiments only after further surface doping of the BL.

\section{Experimental and Theoretical Techniques}

Scanning tunneling microcopy (STM) and spectroscopy (STS) experiments were performed in a multi-chamber ultrahigh-vacuum (UHV) system with a base pressure below $1 \cdot10^{-10}\,\mathrm{mbar}$ using a home-built variable temperature STM similar to the one described in Ref. [\onlinecite{Kuck:2008}]. Both the tip, electrochemically etched from polycrystalline W wire, and sample were cooled by a continuous flow He cryostat to $T = 30\,\mathrm{K}$. Constant current STM images were taken in constant current mode at a tunneling current $I$ with the bias voltage $V$ applied to the sample. Maps of the differential conductance $\mathrm{d}I/\mathrm{d}V$, called STS images in the following, were measured with closed feedback in constant current mode via lock-in technique with a small modulation voltage $V_\mathrm{mod}=10\,\mathrm{mV}$ added to $V$. Such maps are closely related to the local electron density of states (LDOS) of the surface at an energy $eV$ with respect to the Fermi energy $E_{\mathrm F}$. The calculation of the FT-STS images is described below.

ARPES measurements of the band dispersion of the sample were performed on the SGM-3 beamline of the ASTRID synchrotron radiation facility~\cite{Hoffmann:2004}. The ARPES spectra have been acquired using a photon energy of $18\,\mathrm{eV}$ and the temperature of the sample was kept at $T \approx 70\,\mathrm{K}$. The energy and angular resolution for ARPES measurements were better than $20\,\mathrm{meV}$ and $0.2\,^\circ$, respectively. 

The stoichiometric Bi$_2$Se$_3$ single crystal used as substrate was grown and characterized as described in Ref. [\onlinecite{Bianchi:2010b}], and is highly $n$-doped resulting from bulk defects. The crystalk was cleaved by the scotch tape method inside UHV at room temperature resulting in a termination by an intact quintuple layer~\cite{dosReis:2013}. Within 15 min after cleavage, the substrate was cooled down to $T = 250\,\mathrm{K}$, and Bi was deposited at a rate of $1\,\mathrm{BL/5 min}$ from a crucible with high purity material using an electron beam evaporator. The growth temperature was optimized for a smooth growth of a closed (111) BL. The coverage of Bi was calibrated using STM images of well-known (111) BL islands on Bi$_2$Te$_3$~\cite{Yang:2012} and is given in fractions of a (111) BL in the following. For the ARPES measurements, the samples were annealed after the growth at $T = 450\,\mathrm{K}$ for 10 min in order to increase the size of the BL islands in these samples.

The DFT calculations were performed using the full-potential linearized augmented plane wave method as implemented in the FLEUR code~\cite{fleur}. DFT is employed in the generalized gradient approximation as given by Ref.~[\onlinecite{Perdew:1996}] including spin-orbit coupling self-consistently. Using the experimental \bise~lattice parameters, the Bi BL was assumed to sit epitaxially on one side of six quintuple-layers of the substrate and the atomic positions of the BL and the first substrate layers were relaxed. In order to match the experimental dispersions, the distance between substrate and BL had to be further increased, as described in Sec.~\ref{sec:DFT_spinpol}.

\begin{figure}[htbp]
	\centering
		\includegraphics[width=\columnwidth]{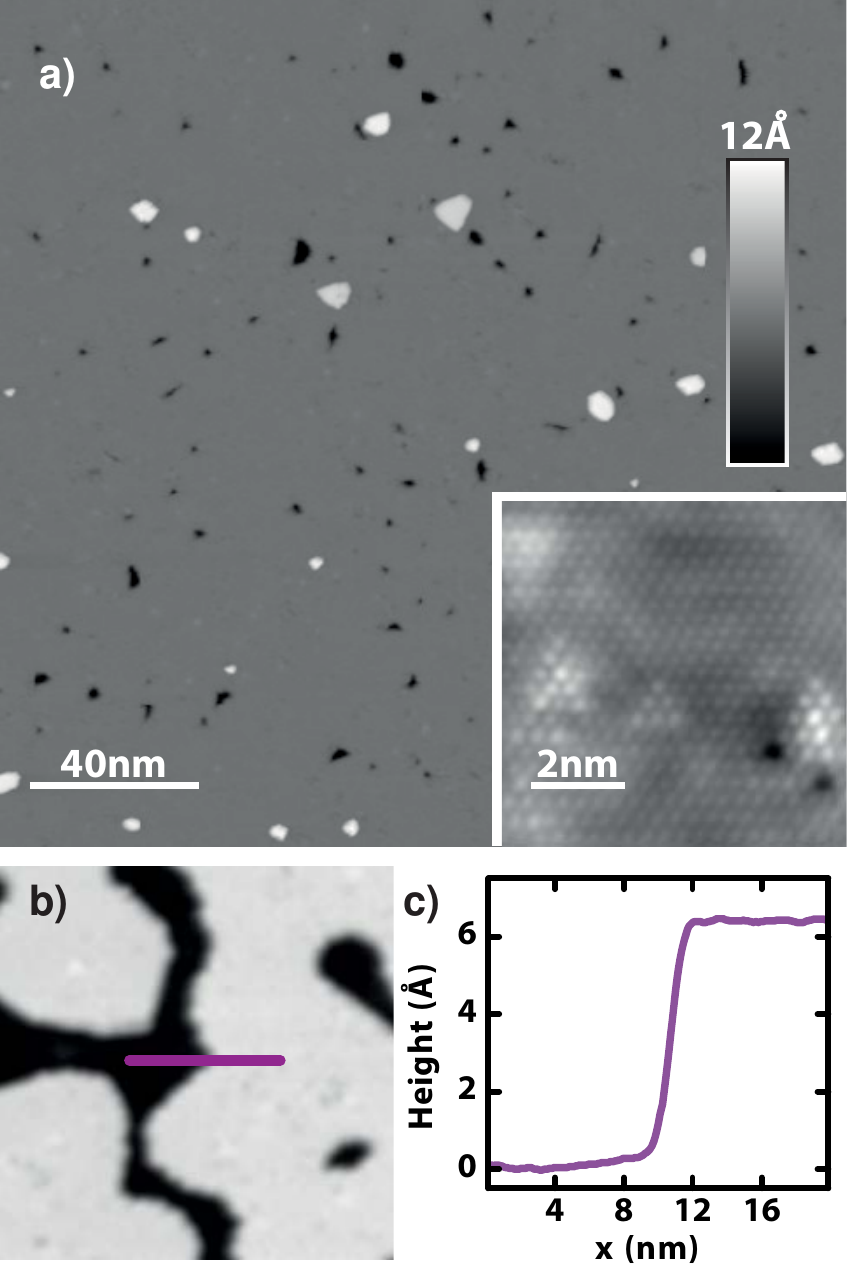}
	\caption{(Color online) Constant current STM image of the Bi BL. (a) Bi BL grown at full coverage on Bi$_2$Se$_3$ ($V=1\,\mathrm{V}$, $I=20\,\mathrm{pA}$). Inset: Atomically resolved STM image taken on a flat area of the BL ($V=-100\,\mathrm{mV}$, $I=5\,\mathrm{nA}$). (b) Bi BL islands grown by deposition of 60\% BL ($V=1\,\mathrm{V}$, $I=20\,\mathrm{pA}$). (c) Height profile taken along the line in (b).
	}
	\label{fig:Figure1}
\end{figure}

\section{Growth and Morphology of the Bi BL on \bise}
\label{sec:STM}
Figure~\ref{fig:Figure1}~(a) shows an STM image of an almost perfectly closed Bi BL on Bi$_2$Se$_3$. There are only a few remaining vacancy islands (dark spots) where the Bi$_2$Se$_3$ substrate is exposed, and a few islands of excess Bi grown on top of the first BL (bright spots). Atomically resolved STM images (inset) reveal an in-plane lattice constant of the BL of $4.1\,\mathrm{\AA}$ which is almost pseudomorphic to the Bi$_2$Se$_3$ lattice, resulting in a lateral compression of the BL of about 10\% with respect to the bulk Bi lattice constant. Deposition of less than 1 BL results in the growth of irregularly shaped Bi BL islands on the bare substrate [Fig.~\ref{fig:Figure1}~(b)]. Note, that, unlike for the case of Bi$_2$Te$_3$~\cite{Yang:2012}, it is not possible to grow triangular shaped BL islands, even if the growth temperature is reduced, indicating a weak bonding of the BL to the \bise~substrate. STM height profiles [Fig.~\ref{fig:Figure1}~(c)] reveal a surprisingly large height of the BL of at least $6.3\,\mathrm{\AA}$, slightly dependent on the bias voltage. Compared to the height of the Bi BL grown on Si(111) of only $4\,\mathrm{\AA}$~\cite{Nagao:2004} which fits with the bulk Bi lattice constant along the trigonal c axis, this indicates a large van der Waals gap between the BL and the Bi$_2$Se$_3$ substrate of roughly $4.4\,\mathrm{\AA}$.

There are only few defects at the surface of the BL which appear as atomic size depressions [see inset of Fig.~\ref{fig:Figure1}~(a)]. Instead, the visible extended triangular shaped protrusions are due to subsurface defects, most probably located at the interface between the BL and Bi$_2$Se$_3$~\cite{Urazhdin:2002, Alpichshev:2012}.

\section{Fourier-Transform Scanning Tunneling Spectroscopy (FT-STS) Results}
\begin{figure*}[htbp]
	\centering
		\includegraphics[width=2\columnwidth]{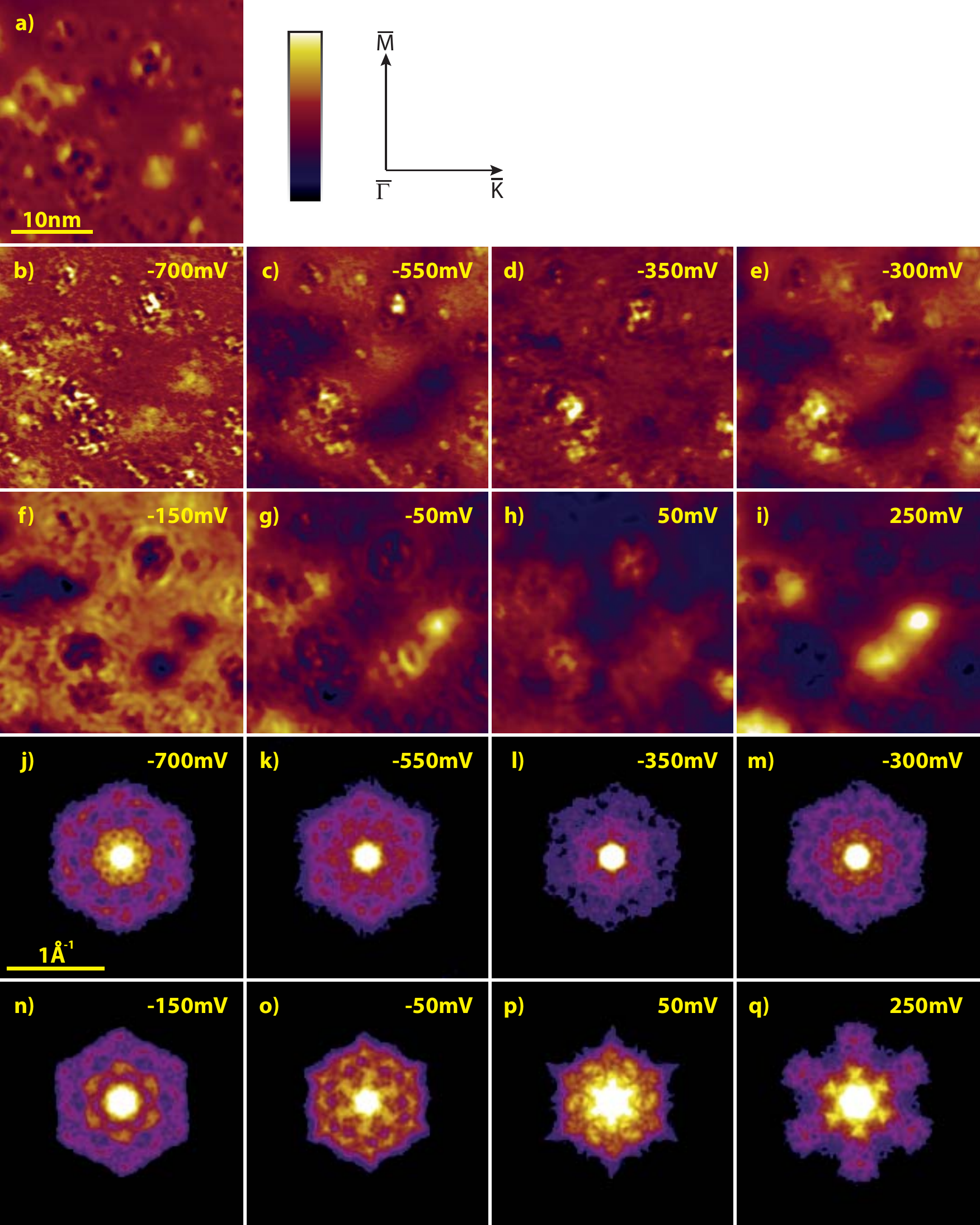}
	\caption{(Color online) (a) Constant current STM image of an area of the closed BL without vacancy islands and Bi excess islands ($V=-50\,\mathrm{mV}$, $I=3\,\mathrm{nA}$, scale bar from 0 to $80~$pm). The crystallographic directions as derived from the atomic resolution in the inset of Fig.~\ref{fig:Figure1}~(a) are indicated. (b-i) STS images of the same sample area as in (a) measured at the indicated sample biases ($I=3\,\mathrm{nA}$). The color range covers the following ranges of conductances: (b) $0.09\,\mathrm{nS}$ to $0.13\,\mathrm{nS}$, (c) $0.09\,\mathrm{nS}$ to $0.17\,\mathrm{nS}$, (d) $0.17\,\mathrm{nS}$ to $0.29\,\mathrm{nS}$, (e) $0.23\,\mathrm{nS}$ to $0.4\,\mathrm{nS}$, (f) $0.4\,\mathrm{nS}$ to $0.7\,\mathrm{nS}$, (g) $1.2\,\mathrm{nS}$ to $2\,\mathrm{nS}$, (h) $0.46\,\mathrm{nS}$ to $1.8\,\mathrm{nS}$, (i) $0.23\,\mathrm{nS}$ to $0.79\,\mathrm{nS}$. (j-q) FT-STS images resulting from Fourier transformation of (b-i) and image processing as described in the text.
	}
	\label{fig:Figure2}
\end{figure*}

In order to study the electronic structure of the BL on Bi$_2$Se$_3$ system, STS images have been taken on a $30\,\mathrm{nm} \times 30\,\mathrm{nm}$ area of the closed BL without vacancy and second BL islands (Fig.~\ref{fig:Figure2}). As visible already in the STM image (a), there are QPI patterns centered mostly around the surface defects, which result from the scattering of the electrons at these defects. The dispersion of the QPI patterns visible in the STS images taken as a function of bias voltage $V$ (b-i) shows a complex behaviour indicating the contribution of more than one electron band. In addition to the rather weak $\mathrm{d}I/\mathrm{d}V$ corrugation due to QPI, there is a rather strong long range contrast which inverts between $-700\,\mathrm{mV}$ and $-550\,\mathrm{mV}$, and again between $-150\,\mathrm{mV}$ and $-50\,\mathrm{mV}$. This contrast reveals potential disorder on a length scale of $10\,\mathrm{nm}$ which is probably caused by donor- or acceptor like subsurface defects~\cite{Morgenstern:2002, Wiebe:2003}.

In order to deduce the origin of the electrons contributing to the QPI, the 2D fast Fourier-transformations of the STS images in Fig.~\ref{fig:Figure2}~(b-i) have been calculated, with the convention of $\left|\textbf{q}\right|=2\pi/\lambda$, resulting in so called FT-STS images.
In such images taken at a bias voltage $V$, the intensity at a given $\textbf{q}$ is related to the probability of an electron scattering from an initial state $\left |E,\textbf{k}\right\rangle$ into a final state $\left |E,\textbf{k'}\right\rangle$ with energy $E=eV$ and wave vectors $\textbf{k}$ and $\textbf{k'}$, respectively, where $\textbf{q}=\textbf{k'}-\textbf{k}$ is the scattering vector~\cite{Simon2011}. The following commonly used~\cite{Chuang:2010, Haenke2012, Okada:2011} image processing steps have been done, in order to reduce the noise in the FT-STS images: First, the STS images have been smoothened by using a median filter with 5 pixel averaging along the slow scan direction. Second, in order to suppress effects due to the infinitely sharp edges of the STS image on the FT, the images have been processed using a window filter with a smooth transition to zero intensity at the rim on a length of about $5\,\mathrm{nm}$. Third, the calculated FT-STS image is symmetrized accoring to the pseudo sixfold symmetry of the underlying bandstructure.
Finally, the FT-STS images have been smoothened using a gaussian filter.

The FT-STS images resulting from the STS images in Fig.~\ref{fig:Figure2}~(b-i) are shown in (j-q). Upon the first inspection, there are two dominant features: below the Fermi energy $E_\mathrm{F}$ the FT-STS signal has a hexagonal rim with the flat side along \gb-\mb, which disperses to smaller $\left|\textbf{q}\right|$ values for increasing energy. Above $E_\mathrm{F}$, the hexagon gets increasingly warped and finally transforms into a star-like shape with the six arms along \gb-\mb. Upon further inspection, there are additional features of smaller size inside the outer hexagon and star, which also disperse as a function of energy, indicating the contribution of two to three bands to the observed QPI throughout the energy window. We took particular care, that these features we analyzed in the following, were not artificially introduced by the image processing steps described above, but are already visible in the raw 2D fast Fourier-transformations of the raw STS images. Moreover, we observed the same energetic evolution of FT-STS features on five sample areas from different cleaves. The comparison of these features from FT-STS data taken on different sample locations showed some energetic shift of at most $200\,\mathrm{mV}$, which we assign to a slight lateral variation of the doping by surface defects. However, the data set we analyzed in the remaining part of this manuscript was taken from one sample location and didn't show such a variation.

Due to the contribution of many bands to the observed QPI, it is a highly nontrivial task to deduce the bandstructure $E(\textbf{k})$ from the distribution of $\textbf{q}$-vectors in the FT-STS-images. On the other hand, by comparison of the $\textbf{q}$-vectors from FT-STS images to $E(\textbf{k})$ determined from complementary ARPES measurements or DFT calculations, it is possible to relate the presence or absence of $\textbf{q}$-vectors in FT-STS images to the scattering-electron spin dependent matrix elements. For this purpose, the FT-STS images are simulated from a DFT calculated bandstructure and compared to the measured FT-STS images in the following.

\section{Simulation of FT-STS images}

\begin{figure*}[htbp]
	\centering
		\includegraphics[width=2\columnwidth]{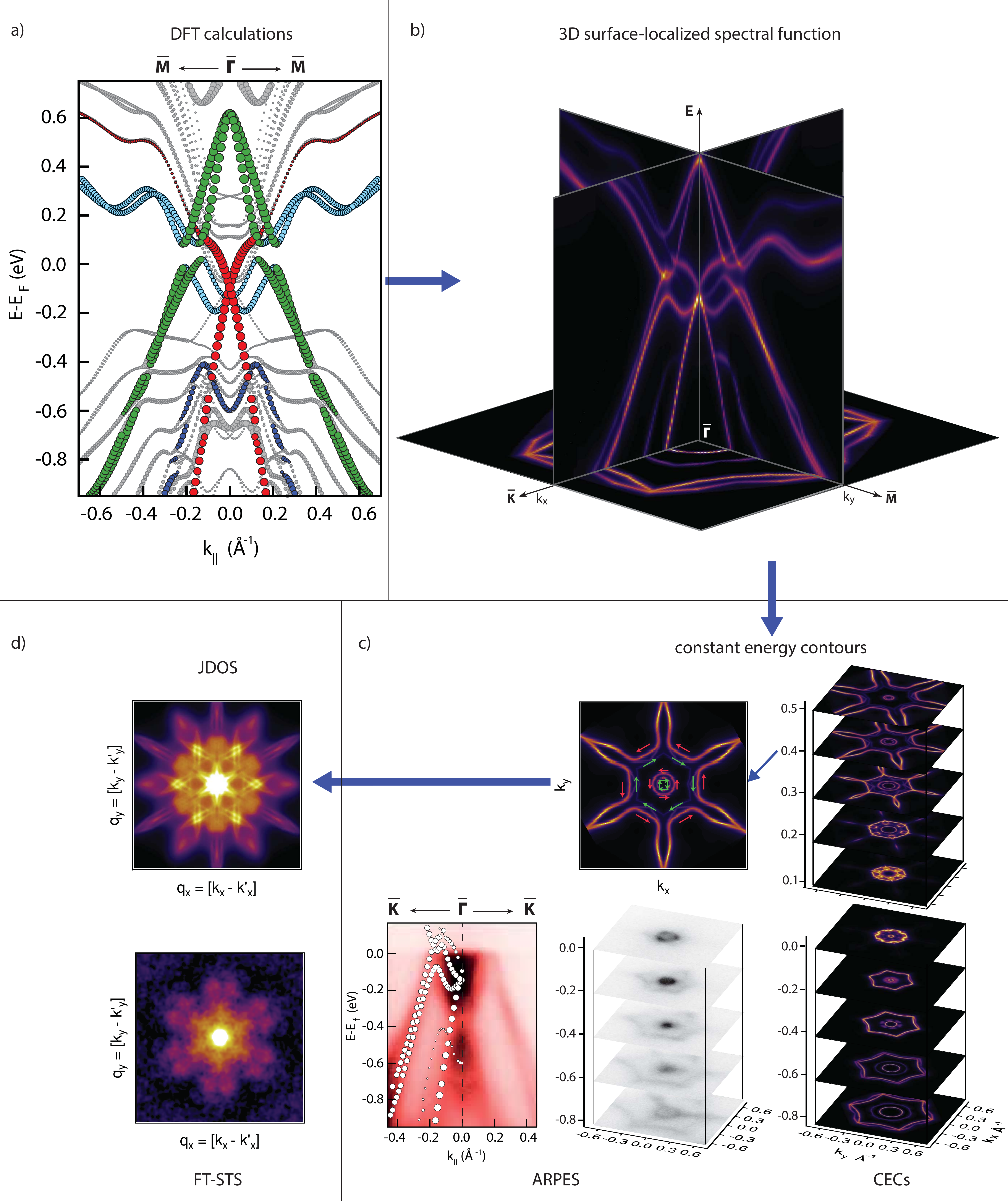}
	\caption{(Color online) (a) DFT calculation of the bands along $\bar{\Gamma}-\bar{M}$. The size of the symbols is proportional to the localization of the state at the surface of the BL. The colors indicate the origin of the four principal bands dominating the surface layer: light blue, Bi$_2$Se$_3$ conduction band (CB); dark blue, Bi$_2$Se$_3$ valence band (VB); green, Bi BL VB; red: topological state (TS). The grey symbols indicate bulk-states or states from the opposite surface of the slab without Bi BL. (b) Three dimensional surface-localized spectral function modelled by 3D-interpolation of the DFT calculated bands of (a) and a similar DFT-calculation of the bands along the other two high symmetry directions in the SBZ ($\bar{\Gamma}-\bar{K}$, $\bar{\Gamma}-\bar{B}$). (c) Right panel: constant energy countours (CECs) through the surface-localized spectral function of (b) at the energies indicated by the vertical axis in electron volts. Top left panel: CEC at $E=0.4\,\mathrm{eV}$ containing the orientations of the electron spin in each band. Bottom left panel: DFT calculation of the bands along $\bar{\Gamma}-\bar{K}$ (dots, the data at positive k is mirror symmetric to the shown data at negative k) superimposed on the band dispersion as measured by ARPES along the same direction (red). Bottom center panel: CECs measured by ARPES at energies in electron volts given by the vertical axis. (d) Comparison between calculated spin-dependent JDOS and measured FT-STS at $E=0.4\,\mathrm{eV}$ ($I=3\,\mathrm{nA}$). $k_x$ is along $\bar{\Gamma}-\bar{K}$ and $k_y$ along $\bar{\Gamma}-\bar{M}$ in (b-d).
	}
	\label{fig:Figure3}
\end{figure*}


The FT-STS images are simulated using the JDOS approximation. \cite{Simon2011} This approach entails computing the self-correlation function of the 2D constant energy contour (CEC) of the surface-localized spectral function $A\left(E,\textbf{k}\right)$ in $\textbf{k}$-space at a given energy $E$:

\begin{equation}
\mathrm{JDOS}(E,\textbf{q})=\int A(E,\textbf{k}) \,M(\textbf{k},\textbf{k}+\textbf{q})\, A(E,\textbf{k}+\textbf{q})d^2\textbf{k}
\label{eq:2}
\end{equation}

Thereby, the number of all possible $\textbf{q}$ within the same CEC is counted and weighted by the spin dependent scattering matrix element $M\left(\textbf{k},\textbf{k'}\right)$. In order to have access to $A$ we develop a semi-theoretical approach, as illustrated in Fig.~\ref{fig:Figure3}. First, the band structure is calculated from an LDA slab calculation of the Bi BL/Bi$_2$Se$_3$ system [Fig.~\ref{fig:Figure3}~(a)], where the slab has been adjusted to the geometry observed by STM and to the Bi BL bands below $E_\mathrm{F}$ as measured by ARPES [Fig.~\ref{fig:Figure3}~(c), see Sec.~\ref{sec:DFT_spinpol}]. The band structure is computed around \gb \hspace{1 pt} along three different directions in a 0.7~$\AA^{-1}$ range in the surface Brillouin zone (SBZ) ($\bar{\Gamma}$-$\bar{K}$, $\bar{\Gamma}$-$\bar{M}$ and $\bar{\Gamma}$-$\bar{B}$, where $\bar{B}$ is defined as the point along the perimeter of the SBZ that sits at equal distance between \kb~and \mb).

Second, the relevant eigenvalues $\epsilon(\textbf{k})$ for each band have been isolated and symmetrized in \textbf{k}-space according to the pseudo sixfold symmetry of the system, which resulted in the description of the band structure along 12 different directions. A 3D interpolation of the data was used to obtain a full space profile of the single electron bare energy $\epsilon(\textbf{k})$. Note that this method does not require fitting the dispersion profile with any function but uses a Voronoi natural neighbors interpolation of close scattered data points. A full 3D spectral function in the energy range between $-1~eV$ and $+0.65~eV$ is then produced according to
\begin{equation}
A(E,\textbf{k})=\frac{|\Sigma''(E,\textbf{k})|}{[E-\epsilon(\textbf{k})-\Sigma'(E,\textbf{k})]^2+\Sigma''(E,\textbf{k})^2}*W(E,\textbf{k})
\label{eq:3}
\end{equation}
where $\Sigma'$ and $\Sigma''$ are real and imaginary parts of the electron self-energy, here set to $0$ and $0.001~eV$, respectively. $W(E,\textbf{k})$ is a weighting factor accounting for different localization of the electron wave function of each state $\left|E,\textbf{k}\right\rangle$ in the Bi BL. In this way it is possible to take care for the surface sensitivity of the STS technique by neglecting all the bands which have low contribution in the BL accoording to DFT, and will therefore not contribute to the QPI pattern. $W$ is defined in all 3D grid points through the same symmetrization-interpolation procedure used for $\epsilon(\textbf{k})$ starting from values associated to the individual points from the DFT calculation [see size of points in Fig.~\ref{fig:Figure3}~(a)]. Figure~ \ref{fig:Figure3}~(b) shows a 3D perspective of the resulting surface-localized spectral function.

Third, we cut $A\left(E,\textbf{k}\right)$ at different $E$ to obtain the CECs as shown in Fig.~\ref{fig:Figure3}~(c). At this point, a good agreement between the simulated CECs and the CECs as measured by ARPES (same figure) demonstrates the soundness of the method.

Finally, the JDOS simulation is performed as described in Equ.~\ref{eq:2} assuming a spin-dependent $M$ (see Sec.~\ref{sec:DFT_spinpol}). The JDOS can be directly compared to the FT-STS data [Fig.~\ref{fig:Figure3}~(d)]. Before this comparison is analysed in detail, the results of the DFT calculations are described in the following.

\subsection{DFT calculated band structure and spin-polarization}
\label{sec:DFT_spinpol}

The slab used for the DFT calculations has been adjusted to the geometry as measured by STM (Sec.~\ref{sec:STM}). The in-plane lattice constant has been set to that of \bise~consistent with the pseudomorphic growth. While the distance between the Bi layers in the BL was kept constant at the bulk Bi value of $1.89\,\mathrm{\AA}$, the distance between the topmost Se layer and the topmost Bi layer in the BL (surface layer) was varied between $4\,\mathrm{\AA}$ and $5\,\mathrm{\AA}$
by increasing the gap between the BL and the substrate, in order to find the best match between the calculated band structure and the ARPES data. This is achieved for $5\,\mathrm{\AA}$, consistent with the large distance between BL and substrate observed by STM (Sec.~\ref{sec:STM}).
The resulting band structure shown in Fig.~\ref{fig:Figure3}~(a,c) reveals four principal bands, which dominate the surface layers by their relatively strong surface localization: The green band can be identified with the Bi BL valence band (VB), which is shifted by about $0.6~eV$ up in energy with respect to the free Bi BL~\cite{Hirahara:2011} due to charge donation to the substrate. A similar charge transfer effect was already proposed for the Bi BL grown on \bite~and was attributed to the difference in work function between the film and the substrate~\cite{Chen:2012d}. Consequently, there is a strong electric field in the first layers of the substrate leading to a downwards band-bending induced quantum well hosting a two-dimensional electron system (2DES) of the \bise~CB in light blue and VB in dark blue, which reveal a similar shape as the respective bands observed for \bise~covered with dilute amounts of adsorbates~\cite{Bianchi:2010b,Bianchi:2011,Benia:2011, Scholz:2012, Loeptien:2014}. Moreover, the electric field causes a Rashba-like spin-splitting as revealed by the two branches of each of the three bands. The band crossing at \gb~of the two spin-split branches of the \bise~CB 2DES was previously named substrate Dirac point (DP)~\cite{Miao:2013c}. However, its origin is of Rashba-type spin-orbit splitting and not of topological nature.

In addition to these bands derived from the Bi BL or from quantum well states of the \bise~substrate, there is a non-trivial 2D TS with a Dirac-like dispersion (red), whose DP appears about $0.1~eV$ below $E_\mathrm{F}$. This TS is strongly localized in the Bi BL and partially in the first quintuple layer of the substrate (see Fig.~\ref{fig:Figure3}~(a) and Ref.~[\onlinecite{Hirahara:2011}]). It {\it replaces} the original 2D TS of the substrate \bise, now coexisting with the Rashba-split \bise~CB 2DES. Hybridization between the Bi BL VB and the \bise~CB 2DES causes a gap in these two bands at about $E = 50\,\mathrm{meV}$ above $E_\mathrm{F}$ which is crossed by the TS. Note, that this hybridization gap is only along \gb--\mb, and not along \gb--\kb~[Fig.~\ref{fig:Figure3}~(b,c)]. Therefore, the necessary condition of an energy gap in the Bi BL bands for the formation of the BL edge state is not fulfilled in this hybridization gap region, as opposed to the argumentation in Ref.~[\onlinecite{Kim:2014}]. In contrast, the edge state is expected to be positioned in the original band gap between VB and CB of the BL at about $E = 0.65\,\mathrm{eV}$ above $E_\mathrm{F}$.

\begin{figure*}[htbp]
	\centering
		\includegraphics[width=2\columnwidth]{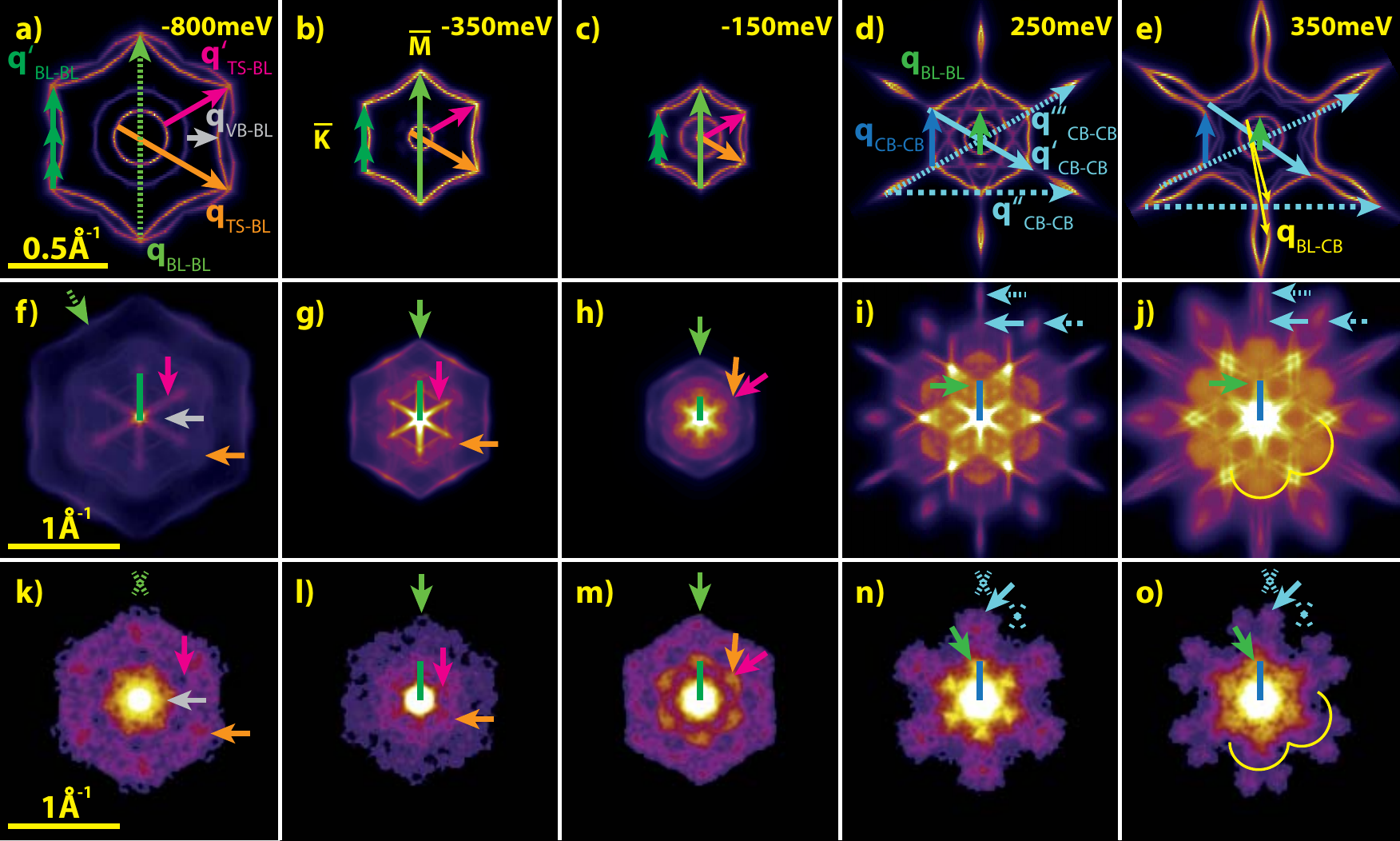}
	\caption{(Color online) Comparison between calculated CECs (a-e), resultant JDOS (f-j), and measured FT-STS (k-o) at the indicated energies. The arrows in the CECs represent scattering vectors $\textbf{q}$, and the arrows or crosses of the same colour indicate the resulting features present or absent, respectively, in the JDOS or FT-STS images.
	}
	\label{fig:Figure4}
\end{figure*}

In order to take into account the spin texture of the Rashba-split bands and of the TS, the spin-dependence of the matrix elements $M$ is considered to lowest approximation, i.e. neglecting possible out-of plane components of the spin and spin-orbit scattering involving changes in the orbital moments \cite{El-Kareh:2014} Out-of plane spin components derive from third order Rashba spin-orbit splitting. Considering the very low anisotropic nature of the split these components are thus expected to be small. Representing the spin part of the electron wave functions by normalized spinors $\left| \textbf{S}_\textbf{k}\right> = \frac{1}{\sqrt{2}} \left( 1,\pm i e^{i \theta_\textbf{k}}\right)$, the matrix elements are calculated as overlap probabilities
\begin{equation}
M\left(\textbf{k},\textbf{k'}\right) = \left| \left< \textbf{S}_\textbf{k} | \textbf{S}_\textbf{k'}\right> \right| ^2 = \frac{1}{2} (1\pm\cos\theta_\textbf{q})
\label{eq:4}
\end{equation}
where $\theta_\textbf{q}$ is the angle between \textbf{k} and \textbf{k'} and $\pm$ takes into account the identity of the two band branches~\cite{Roushan2009,Gomes2009, Wang:2013}. Thereby, the matrix element depends on the projection of the spin in $\left |E,\textbf{k}\right\rangle$ onto the spin in $\left |E,\textbf{k'}\right\rangle$. The spin directions in a CEC are exemplarily represented in Fig.~\ref{fig:Figure3}~(c).

\section{Comparison of FT-STS and JDOS}

Figure~\ref{fig:Figure4} shows a comparison of the simulated CECs, the resulting JDOS images and the measured FT-STS images. First focusing on (a-e) [as well as Fig.~\ref{fig:Figure3}~(c)], the CECs show the following structures: Below $E_\mathrm{F}$ up to about $E = -200\,\mathrm{mV}$, the most dominating features are a central concentric ring stemming from the largely isotropic TS, and two outer hexagonal contours stemming from the spin-split Bi-BL, which has a hexagonal anisotropy reflecting crystal symmetry. In between, there are faint features stemming from the \bise~VB. Note, that these CECs are also confirmed by the ARPES measurements [Fig.~\ref{fig:Figure3}~(c)]. Around $E_\mathrm{F}$, or, more specifically, going from $E = -200\,\mathrm{mV}$ to $E = 100\,\mathrm{mV}$, the TS circle vanishes at the DP and opens again, while the Bi-BL hexagons gradually merge with complex features stemming from the onsetting \bise~CB 2DES. Finally, in the unoccupied DOS above $E = 200\,\mathrm{mV}$, the two inner CECs stem from the spin-split Bi-BL VB, which get more and more isotropic approaching the band maximum. The star-like shape with the arms along \gb--\mb~originates from the \bise~CB which has a strong trigonal anisotropy.

\begin{figure*}[htbp]
	\centering
		\includegraphics[width=2\columnwidth]{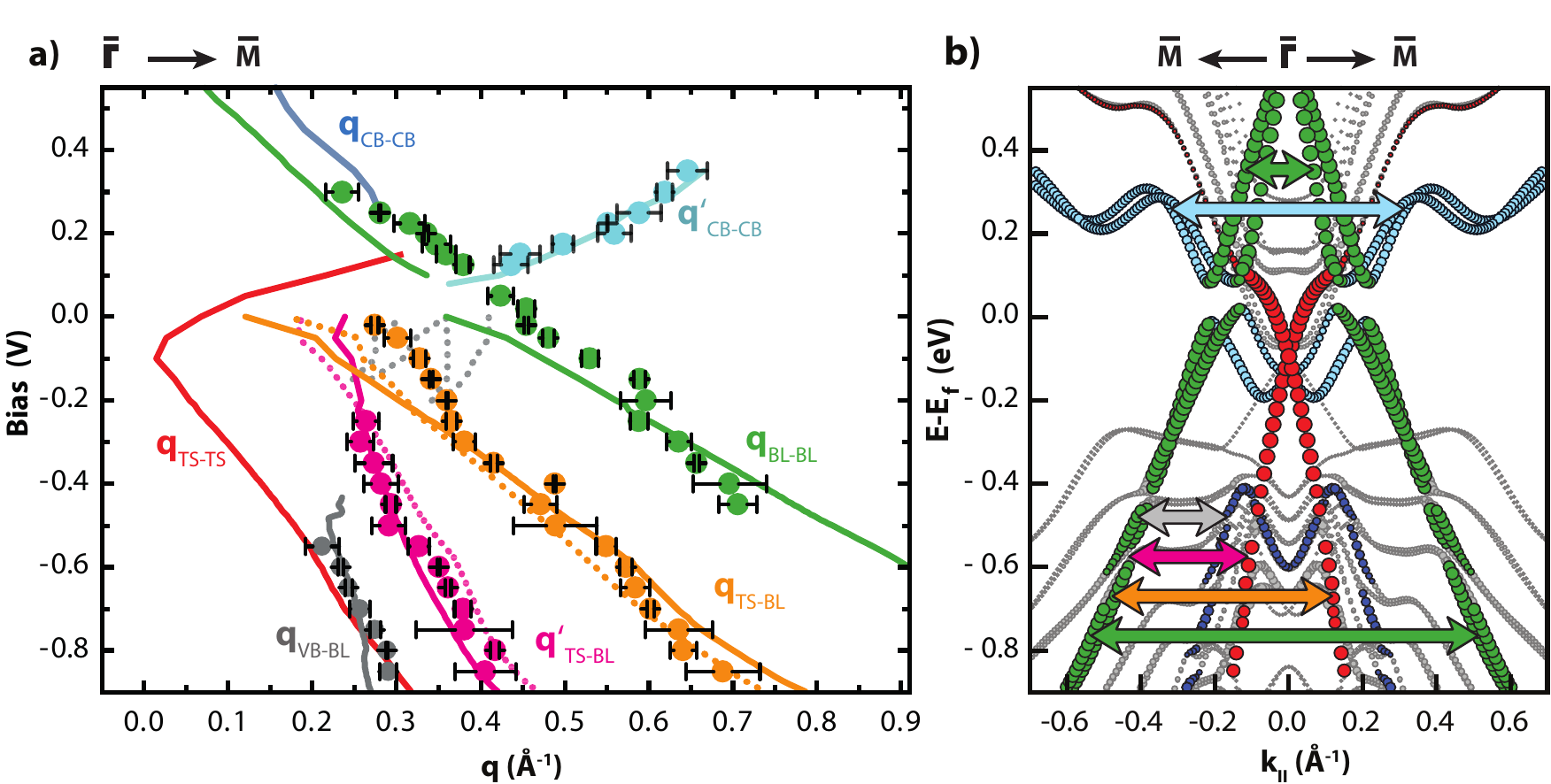}
	\caption{(Color online) (a) Comparison of the dispersion of the scattering vectors from FT-STS (dots) and JDOS simulation (lines) along \gb-\mb~corresponding to the following $\textbf{q}$ as indicated: Bi BL intraband scattering (green), \bise~CB intraband scattering (light blue), interband scattering between Bi BL and \bise~VB (gray), and between Bi BL and TS (pink and orange). The dotted pink and orange lines are the corresponding interband scattering channels of the TS into the second branch of the Bi BL band of equal spin helicity, which are forbidden by spin conservation. Dark blue line: scattering alongside the hexagonal part of the Bi BL CEC [see arrow of same colour in Fig.~\ref{fig:Figure4}~(d,e)]. Grey dotted lines: \bise~CB intraband scattering vectors. The red line indicates the dispersion of the TS intraband scattering which is neither observed in the JDOS, nor in the FT-STS images. (b) DFT calculated band structure including the observed scattering vectors $\textbf{q}$ as arrows with the same colours as the corresponding dispersions in (a).
	}
	\label{fig:Figure5}
\end{figure*}

Taking into account the spin-polarization within these CECs to first approximation (Sec.~\ref{sec:DFT_spinpol}), the JDOS simulation of the QPI results in the images shown in Fig.~\ref{fig:Figure4}~(f-j). The following principal features can be observed and ascribed to particular scattering transitions inside the CECs (in Fig.~\ref{fig:Figure4} the corresponding scattering vectors have been mostly drawn along \gb--\mb, as an example, while the corresponding vectors along \gb--\kb~are left out, in order to increase visibility). Below $E_\mathrm{F}$ [c.p. Figs.~\ref{fig:Figure4}~(a,b,c,f,g,h)] intraband scattering through ($\textbf{q}_\mathrm{BL-BL}$) and along the side ($\textbf{q}^{'}_\mathrm{BL-BL}$) of the almost degenerate spin-split hexagonal Bi BL CECs produces the inner star (green line) and outer hexagonal rim (green arrow) in the JDOS, respectively. Interband scattering between the TS and the Bi BL ($\textbf{q}_\mathrm{TS-BL}$ and $\textbf{q}^{'}_\mathrm{TS-BL}$) results in two additional inner hexagonal features (orange and pink arrows) in the JDOS. Below the onset of the \bise~VB, there is an additional interband scattering channel ($\textbf{q}_\mathrm{VB-BL}$) from the \bise~VB into the Bi-BL VB [grey arrow in (a)], which is not clearly visible in the JDOS (f). Above $E_\mathrm{F}$ [c.p. Figs.~\ref{fig:Figure4}~(d,e,i,j,)] the intraband Bi BL scattering between the two branches of opposite spin chirality ($\textbf{q}_\mathrm{BL-BL}$) and the intraband scattering $\textbf{q}_\mathrm{CB-CB}$ along the flat side of the \bise~CB CEC have almost equal scattering vectors and lead to the inner star shaped feature in the JDOS (green arrow and dark blue line). The different intraband scattering vectors involving the six arms of the star-shaped \bise~CB CEC ($\textbf{q}^{'}_\mathrm{CB-CB}$, $\textbf{q}^{''}_\mathrm{CB-CB}$, $\textbf{q}^{'''}_\mathrm{CB-CB}$) lead to the outer star shape of the JDOS with 12 arms. The additional interband scattering vectors (not shown) lead to very complex features in the $\textbf{q}$-space in between these two intraband scattering features.

Comparison of these simulated JDOS images and the measured FT-STS images [c.p. Figs.~\ref{fig:Figure4}~(f-j) to (k-o)] leads to the conclusion that the overall trend in the experimentally observed QPI patterns can be understood as follows. While below $E_\mathrm{F}$ the dominant hexagonal outer shape in the FT-STS images results from intraband scattering $\textbf{q}_\mathrm{BL-BL}$ in the Bi BL, the dominant star-like shape above $E_\mathrm{F}$ comes from one of the intraband scattering vectors, i.e., $\textbf{q}^{'}_\mathrm{CB-CB}$. Moreover, by thorough comparison almost all of the possible interband and intraband scattering vectors found in the simulated JDOS images can be identified in the measured FT-STS images [see coloured arrows in (f-j) and (k-o)]. Most importantly, the interband scattering between the TS and the Bi-BL, $\textbf{q}_\mathrm{TS-BL}$ and $\textbf{q}^{'}_\mathrm{TS-BL}$, are observed over the entire energy range below $E_\mathrm{F}$ down to almost $E =-1\,\mathrm{eV}$. Another important feature visible in the JDOS and FT-STS images is the flower-like pattern marked in Figs.~\ref{fig:Figure4}~(j,o) in yellow. This pattern stems from scattering vectors $\textbf{q}_\mathrm{BL-CB}$ between the star-shaped CEC of the \bise~CB and the circular CEC of the Bi BL [Fig.~\ref{fig:Figure4}~(e)] and proves the coexistance of these two bands in the unoccupied states region.

Some of the scattering vectors identified in the JDOS are missing in the FT-STS images. The $\textbf{q}_\mathrm{BL-BL}$ scattering channel is absent in the FT-STS images at very large negative bias voltage [green cross in (k)]. Also, from the three intraband \bise~CB scattering vectors dominating the JDOS, only $\textbf{q}^{'}_\mathrm{CB-CB}$ is visible in the FT-STS [light blue arrow and crosses in (o)]. These discrepancies are probably explained by deficiencies in the matrix elements $M$, as will be discussed in Sec.~\ref{sec:discussion}. As will be shown in the following section, a comparison of the dispersions of the $\textbf{q}$ vectors, which are independent of scattering matrix element effects, reveals an astonishingly good agreement between measurement and simulation.

\section{Comparison of measured and simulated $\textbf{q}$ dispersion}

The length of the scattering vectors ($\left|\textbf{q}\right|$) along \gb--\mb~of all experimentally detected QPIs has been extracted from the FT-STS images and is plotted in Fig.~\ref{fig:Figure5}~(a) as a function of energy (dots) together with the dispersion extracted from the spectral function and JDOS images (lines). For comparison, the DFT-calculated band structure which is the basis for the simulation, is plotted in the same figure together with the according scattering vectors (b). There is an excellent overall agreement between the measured and simulated dispersions over most of the energy range. Note, that $\textbf{q}_\mathrm{TS-TS}$ is plotted only for reference in Fig.~\ref{fig:Figure5}~(a), but detected neither in the JDOS nor in the FT-STS images, satisfying the avoided backscattering as a well known property of time-reversed band partners in general and a TS in particular. Also, the BL intraband scattering appears exclusively between the two branches of equal spin helicity [Fig.~\ref{fig:Figure5}~(b)]. Only in a small energy window from $E_\mathrm{F}$ down to roughly $E = -200\,\mathrm{meV}$, there is a considerable deviation for $\textbf{q}_\mathrm{TS-BL}(E)$ and $\textbf{q}_\mathrm{BL-BL}(E)$. This deviation occurs in an energy range, where there are principally a number of additional possible \bise~CB intraband scattering vectors which might overlap with $\textbf{q}_\mathrm{TS-BL}$ (grey dotted lines). Above $E_\mathrm{F}$, $\textbf{q}_\mathrm{BL-BL}$ and $\textbf{q}_\mathrm{CB-CB}$ are almost degenerate, such that the experimentally detected dispersion is probably a mixture of these two scattering channels. 

\section{Discussion and Conclusions}
\label{sec:discussion}
Taking into account the relative simplicity of the used JDOS method, and the DFT based approach for the spectral function, which typically comprises some error in the energetic positions of the bands around $E_\mathrm{F}$, the correspondence between both QPI interference patterns (Fig.~\ref{fig:Figure4}) and dispersion (Fig.~\ref{fig:Figure5}) in simulation and experiment is astonishingly good. The average value of the cross correlation between the experimental FT-STS and the simulated JDOS images defined by
\begin{equation}
\frac{\sum\limits_{i,j}[(I_{\mathrm{exp}}(i,j)-<I_{\mathrm{exp}}>)\cdot (I_{\mathrm{sim}}(i,j)-<I_{\mathrm{sim}}>)]}{\sqrt{\sum\limits_{i,j}(I_{\mathrm{exp}}(i,j)-<I_{\mathrm{exp}}>)^2}\cdot\sqrt{\sum\limits_{i,j}(I_{\mathrm{sim}}(i,j)-<I_{\mathrm{sim}}>)^2}}
\end{equation}
where $I$ are the intensities at point $(i,j)$ of the images and $<I>$ denote their mean intensity values, amounts to 80\%, with a variation of $\pm 10$\% across the whole investigated energy range.
Figure~\ref{fig:Figure5} nicely illustrates, that the intraband scattering inside the TS is strongly suppressed by time reversal symmetry due to its peculiar spin texture over a large energy range. Nevertheless, {\it interband} scattering between the TS and the Bi BL band is allowed. The approximate linearity of $\textbf{q}_\mathrm{TS-BL}(E)$, $\textbf{q}^{'}_\mathrm{TS-BL}(E)$, and $\textbf{q}_\mathrm{BL-BL}(E)$, taking into account the inversion symmetry of the band structure with respect to \gb, experimentally proves the presence and linear dispersion of the TS. This holds even in the energy range where the TS is almost degenerate with the substrate VB. This extraordinary property of the particular TS in Bi BL/\bise~is caused by its localization in the Bi BL which suppresses hybridization with bulk VB states in contrast to other 3D TIs. However, the observed interband scattering from the TS into the Bi BL VB excludes the possibility of using this TS for carrying dissipationless spin currents on the surface.

On the other hand, there are obvious discrepancies between the FT-STS and JDOS images shown in Fig.~\ref{fig:Figure4}, indicating a too simplistic form of the assumed scattering matrix elements $M$. The absence of $\textbf{q}_\mathrm{BL-BL}$ at low energy [c.p. green arrow and cross in Fig.~\ref{fig:Figure4}~(f,k)] and of the two largest \bise~CB intraband scattering vectors [c.p. light blue dotted and dashed arrows and crosses in Fig.~\ref{fig:Figure4}~(d,e,i,j,n,o)], in the FT-STS images might be explained by the sensitivity of STS to the vacuum density of states whereas, the JDOS approach considers the density of states {\it in} the surface. The vacuum density of states is dominated by states with small $\textbf{k}$, damping QPI patterns involving large $\textbf{q}$.
By switching off the $\textbf{q}_\mathrm{BL-BL}$ scattering channel in the JDOS simulations a very good agreement with the FT-STS image is found. Another explanation for the missing \bise~CB intraband scattering vectors is spin selection, which may be oversimplified in the JDOS approach, in particular for the strongly anisotropic CB where the constraint to in-plane components of the spin breaks down. The inclusion of a more rigorous treatment of the electron spin in the matrix elements is beyond the scope of the present paper.

The experimentally determined band structure substantiates that the original band gap of the Bi BL is moved up to $E \approx0.6\,\mathrm{eV}$ above $E_\mathrm{F}$. This doping effect comes about by the charge transfer from the BL to the substrate, which is also responsible for the Rashba spin-split substrate CB 2DES states visible in the band structure of Fig.~\ref{fig:Figure5}~(b). The strong Rashba splitting has been recently observed for Bi BL grown on a single quintuple layer of \bise (or \bite)~\cite{Wang:2013}, although, there, it does not coexist with a TS. It remains to be experimentally studied, whether the energetic position of the topological edge state of the BL is in the BL band gap, as predicted theoretically~\cite{Yang:2012}, and thus not accessible to transport experiments. In this case, it might be possible to shift the edge state to $E_\mathrm{F}$ by surface doping using decoration of the BL with alkali adatoms~\cite{Bianchi:2012b, Loeptien:2014}. Finally, our paper illustrates a methodology for the analysis of FT-STS images by DFT-based spin-dependent JDOS simulations enabling to pin point the origin of very complex QPI interference patterns which stem from the electrons in several energetically degenerate bands.

\section{Acknowledgements}
We thank Tien-Ming Chuang for his help concerning the fast Fourier transformation procedures, and Lucas Barreto and Wendell Silva for the help with the ARPES data. Computing time on the supercomputer JUROPA at the J{\"u}lich Supercomputing Centre (JSC) is gratefully acknowledged. We acknowledge financial support from the DFG via SPP 1666 "Topological Insulators: Materials - Fundamental Properties - Devices". AE, JW, and RW acknowledge financial support from the DFG via SFB 668. AAK acknowledges Project No. KH324/1-1 from the Emmy-Noether-Program of the DFG. MM, XGZ, and PhH acknowledge financial support from the VILLUM foundation and from the Aarhus University Research Foundation. JLM and BBI acknowledge the Danish National Research Foundation (DNRF93).

%


\end{document}